\documentclass[twocolumn,nofootinbib,superscriptaddress]{revtex4}

\usepackage{graphicx}
\usepackage{dcolumn}
\usepackage{bm}
\usepackage{hyperref}
\usepackage{textcomp}
\usepackage{amsmath}
\usepackage{multirow}

\begin{document}
\title{The Physics Potential of the LENA Detector
\footnote{presented at Cracow Epiphany Conference, 5-8 January 2010}%
}
\author{M. Wurm}\email[Corresponding author, e-mail:~]{mwurm@ph.tum.de}
\author{F. von Feilitzsch}
\author{M. G\"oger-Neff}
\author{T. Lachenmaier}
\author{T. Lewke}
\author{Q. Meindl}
\author{R. M\"ollenberg}
\author{L. Oberauer}
\author{J. Peltoniemi}
\author{W. Potzel}
\author{M. Tippmann}
\author{J. Winter}
\affiliation{Physik-Department E15, Technische Universit\"at M\"unchen, James-Franck-Str., D-85748 Garching, Germany}


\begin{abstract}
\noindent The large-volume liquid-scintillator detector LENA (Low Energy Neutrino Astronomy) has been proposed as a next-generation experiment for low-energy neutrinos. High-precision spectroscopy of solar, Supernova and geo-neutrinos provides a new access to the otherwise unobservable interiors of Earth, Sun and heavy stars. Due to the potent background discrimination, the detection of the Diffuse Supernova Neutrino Background is expected for the first time in LENA. The sensitivity of the proton lifetime for the decay into $K^+\bar\nu$ will be increased by an order of magnitude over existing experimental limits. Recent studies indicate that liquid-scintillator detectors are capable to reconstruct neutrino events even at GeV energies, providing the opportunity to use LENA as far detector in a long-baseline neutrino beam experiment.    
\end{abstract}

\maketitle

  
\section{Introduction}
\noindent The recent achievements of the Borexino \cite{bx08be7,bx10geo} and KamLAND  \cite{kam05,kam08} experiments demonstrate the large potential of liquid-scintillator detectors (LSDs) as low-energy neutrino observatories \cite{bx08det,kam04}. Based on these successes, a large-volume LSD has been proposed from various sides \cite{hsd, lea08}: In Europe, the 50\,kton liquid scintillator detector LENA \cite{obe04,mar06,wur07,mar08} 
is one of the three options discussed for a next-generation neutrino observatory: Together with the water \v{C}erenkov detector MEMPHYS\footnote{MEgaton Mass PHYSics} (500\,kt) \cite{bel06} and the liquid-argon TPC GLACIER\footnote{Giant Liquid Argon Charge Imaging ExpeRiment} (100\,kt) \cite{rub09}, LENA is part of the European LAGUNA\footnote{Large Apparatus for Grand Unification and Neutrino Astrophysics} design study that currently investigates candidate sites for a new laboratory hosting one of the large-scale detectors \cite{lag07,lag-www}.

LENA scales the technology and design of Borexino \cite{bx08det} to a target mass of 50\,ktons, greatly enhancing the detection sensitivity. This provides the opportunity to observe low-energy neutrinos from various astrophysical and terrestrial origins: High-statistics measurements of already established neutrino sources like a galactic Supernova (SN) explosion, the Sun or the Earth's interior will resolve energy spectra and the time development of the $\nu$ signals in unprecedented detail \cite{win07dpl, tod08dpl, hoc05}. At the same time, the search for very rare events becomes possible, as the excellent background rejection allows to identify a handful of events out of several years of data taking. Thus, LENA will be able to observe the faint flux of the Diffuse Supernova Neutrino Background (DSNB) \cite{wur07dsn}. Moreover, an increase in sensitivity for proton decay search is expected, the new lifetime limits for the $K^+\bar\nu$ decay channel surpassing current limits by more than an order of magnitude \cite{mar05}.

Recent studies indicate that a large-volume LSD can resolve both momentum and energy of GeV particles on a level of a few percent \cite{lea09,pel09tra}. Moreover, Monte Carlo (MC) simulations studying the reconstruction capability of a LSD for the complex event topologies of charged-current neutrino interactions show promising results \cite{pel09tra}. These techniques offer the opportunity to complement the rich low-energy physics program of LENA by long-baseline neutrino oscillation experiments, either from atmospheric neutrinos or an accelerator-produced neutrino beam \cite{pel09sup}.

The present contribution provides an overview of the intended detector design (Sect.\,\ref{SecDetDes}) and the phenomenological studies performed for low-energy neutrino sources (Sect.\,\ref{SecLowEne}). Furthermore, it reports on the current status of GeV track reconstruction, and presents the potential of LENA for proton decay search and as a far detector in a long-baseline neutrino beam experiment (Sect.\,\ref{SecHigEne}). Results are summarized in Sect.\,\ref{SecSum}.

\section{Detector Design}
\label{SecDetDes}

\begin{figure}
\centering
\includegraphics[width=0.4\textwidth]{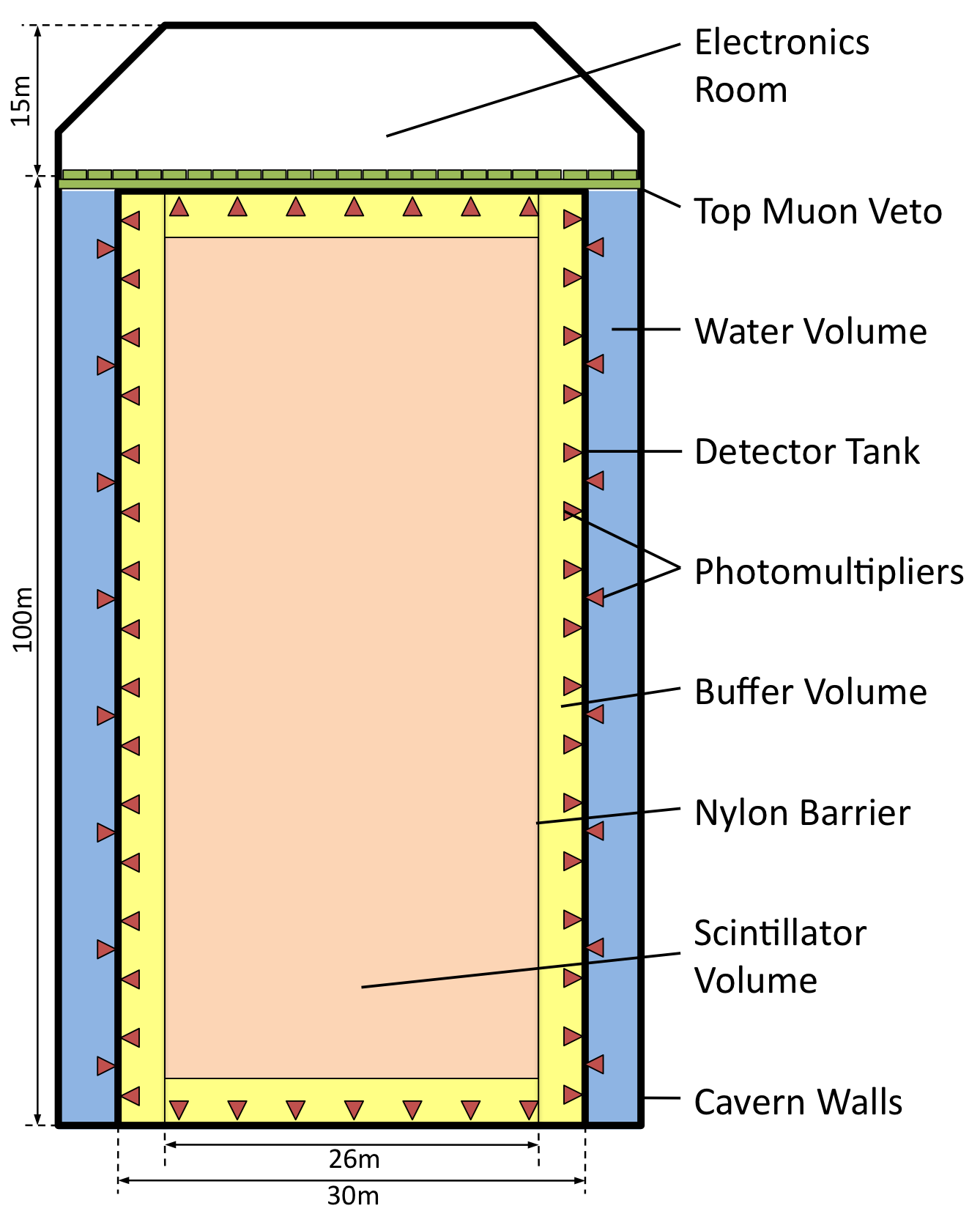}
\caption{Sketch of the LENA detector layout. \label{FigLENA}}
\end{figure}

\noindent The basic design concept of LENA is very similar to those of the Borexino detector \cite{bx08det}: 
As depicted in Fig.\,\ref{FigLENA}, the detector components are arrayed in cylindrical shells around the central scintillator volume (SV) that is unsegmented. Due to the limits set by the optical transparency of the solvent, this volume is cylindric instead of spheric to reduce the maximum light path to the most nearby photomultipliers (PMs): An upright cylinder of 26\,m in diameter and 100\,m in height corresponds to an active volume of 5.3$\times$10$^4$\,m$^3$. As liquid scintillators feature densities of 0.85 to 0.99 t/m$^3$, the target mass ranges from 45 to 53\,kt. Laboratory experiments have shown that phenyl-xylyl-ethane (PXE, with a possible admixture of dodecane) \cite{bx04pxe} and pure linear alkylbenzene (LAB) \cite{SNO08phd} are adequate solvents, providing fast signal decay times and attenuation and scattering lengths exceeding 10\,m at a wavelength of 430\,nm \cite{SNO08phd, mar09, wur10sca}. 

The SV is separated from the surrounding buffer volume (BV) by a cylindrical nylon barrier. This 2\,m thick shell contains an inactive buffer liquid that shields the SV from external radioactivity, allowing to define the bulk of the SV as fiducial volume (FV) for neutrino detection. The liquids in both volumes will be of comparable density to avoid buoyancy effects. SV and BV are contained in a cylindrical steel tank of 100\,m height and 30\,m diameter. Concrete is considered as an alternative tank material. While concrete is preferable from an engineering and monetary point of view, the larger natural radioactivity content requires an increase in tank diameter to obtain a FV of comparable size.

The radially inward-looking PMs are mounted at the steel tank that contains SV and BV. The aimed-for optical coverage of the detector walls is 30\,\%: This requires 13\,500 20-inch PMs of the Super-Kamiokande type. The total photosensitive area can be reduced at least by a factor of 2 by equipping the PMs with reflective Winston cones. These light concentrators increase the photon collection area per PM while limiting the field of view of the PM. The optimum size will be determined in MC simulations of the detector performance. Currently favored are designs using 8'' or 5'' PMs equipped with Winston cones, requiring about 45\,000 or 110\,000 pieces, respectively. The PMs are mounted to a narrow scaffolding in front of the cylinder walls. Cabling for high-voltage supply and signal readout runs behind the PMs inside the tank. High-voltage supply, voltage divider, and signal digitizers are planned to be attached directly to the PM base, improving the signal transmission fidelity of the 100\,m long cables. Several channels might share a single close-by electronics unit both for HV supply and signal read-out, greatly reducing the amount of cabling. This option is currently studied in collaboration with the PMm2 group \cite{gen09}.

The space between the detector tank and the surrounding cavern is filled with water. It provides additional shielding from external radioactivity and fast neutrons. This water volume can also be exploited as a \v{C}erenkov medium for the active veto of cosmic muons, which implies the installation of about 1500 additional PMs on the outer tank wall. The water volume also simplifies construction, especially for the concrete tank, as the pressure of water on the outside of the tank walls stabilizes the whole structure.

To provide a muon veto for the top lid of the detector tank, layers of plastic scintillator panels or resistive plate chambers (RPCs) will cover an area slightly larger than the tank. Especially in a deep mine, most of the cosmic muon flux is vertical: A dense and highly segmented instrumentation on top of the detector will provide high detection efficiency and precise muon tracking, optimizing background rejection.

Shape and lining of the cavern walls as well as the cavern depth are dependent on the detector site. The issue of an adequate underground laboratory is currently investigated in the European LAGUNA site study \cite{lag-www}. Vertically curved cavern walls are the preferred solution to resist the horizontal rock stresses in the bedrock present at the Pyh\"asalmi mine in Finland. Under different geological circumstances that are for instance found at the Fr\'ejus site in France, straight cavern walls covered by a relatively thick layer of concrete are the optimum solution (Fig.\,\ref{FigLENA}). In any case, a minimum width of 2\,m water buffer is set as a design benchmark to provide the necessary shielding from fast neutrons. The designated rock overburden of 4000\,mwe is achievable for both sites.

\section{Low-Energy Neutrino Physics}
\label{SecLowEne}

\noindent LENA is motivated as an observatory for low-energy astrophysical neutrinos: The large volume will allow the detection of solar \cite{tod08dpl}, Supernova (SN) \cite{win07dpl} and geoneutrinos \cite{hoc05} at unprecedented event rates. The expected signal and the scientific gain regarding the understanding of the neutrino sources and the intrinsic parameters of the neutrinos are outlined in this section. In addition, we will highlight the possibility for the first-time detection of the DSNB in LENA \cite{wur07dsn}. 

It has been pointed out that LENA will be sensitive to neutrinos produced by the annihilation of dark matter particles, provided these particles are relatively light \cite{pal07}. Moreover, if built close to one or several strong nuclear power reactors (e.\,g.\,at the Fr\'ejus site), a measurement similar to the KamLAND experiment could be reproduced, reducing the uncertainties of the solar mixing parameters $\theta_{12}$ and $\Delta m_{12}^2$ to 10\,\% and 1\,\% at 3$\sigma$, respectively \cite{pet06}.

\subsection{Galactic Supernova Neutrinos}

\noindent In case of a core-collapse SN within the Milky Way, a huge burst of neutrinos is expected to reach terrestrial detectors. The short neutronization burst of $\nu_e$ due to the conversion of protons to neutrons in the collapsing iron core is followed by an enormous signal of $\nu\bar\nu$ pairs of all flavors lasting for about 10 seconds. The latter are produced in the cooling phase of the developing proto-neutron star, radiating away about 99\,\% of the released gravitational energy \cite{kot05}.

While the dominant $\nu$ detection channel in present-day LSDs is the inverse beta decay (IBD), $\bar\nu_e + p \rightarrow n + e^+$, the target mass of LENA is large enough to exploit a variety of reaction channels for all flavors. In the standard SN scenario that describes the explosion of an 8\,M$_{\odot}$ progenitor star at 10\,kpc distance, LENA will detect between 10\,000 and 15\,000 events \cite{win07dpl}. The numbers vary with the assumed SN neutrino spectra and with the occurrence of matter effects in the stellar envelope \cite{kot05,tot97,tho02,kei02}. An overview of the detection channels and their rates in LENA is given in Tab.\,\ref{TabLnSN}.

\begin{table}
\begin{center}
\begin{tabular}{|cl|c|}
\hline
\multicolumn{2}{|l|}{Channel} & Rate  \\
\hline
(1) & $\bar\nu_{e}+p \rightarrow n + e^+$ 								& 7\,500$-$13\,800 \\
(2) & $\bar\nu_{e}+{^{12}\mathrm{C}}\rightarrow{^{12}\mathrm{B}}+e^+$	& 150$-$610 \\
(3) & $\nu_{e}+{^{12}\mathrm{C}}\rightarrow{^{12}\mathrm{N}}+e^-$		& 200$-$690 \\
(4) & $\nu_{e}+{^{13}\mathrm{C}}\rightarrow{^{13}\mathrm{N}}+e^-$		& $\sim$10 \\
(5) & $\nu+{^{12}\mathrm{C}}\rightarrow{^{12}\mathrm{C}^*}+\nu$			& 680$-$2\,070 \\
(6) & $\nu+e^-\rightarrow e^-+\nu$									& 680 \\
(7) & $\nu+p\rightarrow p+\nu$										& 1\,500$-$5\,700 \\
(8) & $\nu+{^{13}\mathrm{C}}\rightarrow{^{13}\mathrm{C}^*}+\nu$			& $\sim$10 \\
\hline
\end{tabular}
\caption{Overview of the detection channels for SN neutrinos available in LENA. The rates are derived in \cite{win07dpl} and depend on the underlying SN model. Total event rates vary from 10\,000 to 15\,000 events for the standard SN scenario (8\,$M_{\odot}$ progenitor in 10\,kpc distance).}\label{TabLnSN}
\end{center}
\end{table}

More than half of the events are caused by the IBD (1) which allows a precision measurement of the $\bar\nu_{e}$ energy spectrum and the temporal evolution of the $\bar\nu_e$ flux.  The  energy resolution of a large-volume LSD offers the possibility to study the imprints of matter effects in the $\bar\nu_{e}$ spectrum that either result from the transit through the progenitor star envelope or the Earth \cite{dig03}. As the occurrence of these effects is closely linked to the size of the mixing angle $\theta_{13}$ and the neutrino mass hierarchy, SN $\nu$ detection in LENA is also sensitive to these up to now undetermined neutrino parameters \cite{kot05}. Further effects like the recently proposed collective oscillations might also be imprinted on the $\bar\nu_e$ spectrum \cite{das09}.

The charged current (CC) reaction of $\nu_{e}$ on Carbon (3) will be mainly used to determine the $\nu_{e}$ flux. The event signature is hard to discern from the CC reaction of $\bar\nu_{e}$'s (2) that features a final state nucleus of comparable lifetimes and beta endpoint. However, statistical subtraction of the $\bar\nu_{e}$ flux which is determined very accurately by channel (1) can be used to isolate the $\nu_{e}$ signal at a 10\,$\%$ level. The remaining uncertainty is mostly due to the uncertainties of the reaction cross sections \cite{win07dpl}. 

While the channels (1-4) allow to discriminate $\nu_{e}$ and $\bar\nu_{e}$, channels (5-8) are accessible for all neutrinos independent of their flavors or anti-flavors. The neutral current (NC) reactions on Carbon (5+8) are flux measurements only and bear no spectral information. Both elastic electron scattering (6) and proton scattering (7) on the other hand provide spectral data for the combined flux of all flavors. The signal on protons is dominantly caused by $\nu_\mu$ and $\nu_\tau$ (and their anti-flavors) as their expected mean energies are assumed to be larger than for $\nu_e$ in some SN models \cite{tot97,tho02}. Due to the strong dependence of the measured event rate on the mean neutrino energy, proton scattering is very sensitive to the temperature of the SN neutrinosphere.

\subsection{The Diffuse Supernova Neutrino Background}

\noindent All SN explosions in the universe contribute to an $-$ on cosmic scales $-$ constant and isotropic flux of neutrinos, the Diffuse Supernova Neutrino Background (DSNB) \cite{and04}. The expected flux can be computed combining measurements of the redshift-dependent Supernova rate with the energy spectra of SN neutrinos that are derived from MC simulations \cite{tot97,tho02,kei02}. With $\sim$10$^2$ neutrinos per cm$^2$s, the DSNB is about eight orders of magnitude fainter than the terrestrial flux of solar neutrinos.

During proto-neutron star cooling, $\nu\bar\nu$ pairs of all flavors are generated. The most accessible for detection are $\bar\nu_{e}$, as the IBD reaction features the largest cross section at low energies. According to contemporary models, about 4 events per year are expected for Super-Kamiokande \cite{hor08}. Unfortunately, the 2.2\,MeV $\gamma$ quantum emitted by the capture of the final state neutron on Hydrogen is below the instrumental detection threshold of a water \v{C}erenkov detector: As a consequence, $\bar\nu_e$ events are covered by various backgrounds\footnote{Recent R\&D activities aim towards the detection of the neutron by adding Gd to the water \cite{hor08,bea03,sk08ane}.}. In an LSD, the neutron capture $\gamma$ is well visible and offers in coincidence with the positron a very distinct detection signature, allowing to reduce the background considerably. However, present day LSDs lack the target mass necessary to detect the DSNB. LENA might be the first experiment capable of detecting these relic SN neutrinos.

\begin{figure}
\centering
\includegraphics[width=0.44\textwidth]{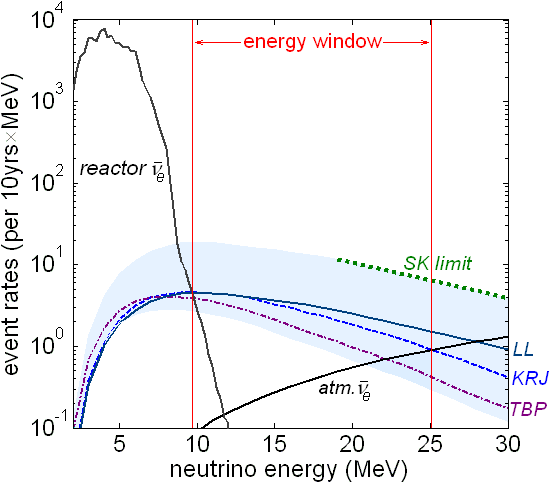}
\caption{Event rates of reactor, atmospheric and DSNB $\bar{\nu}_{e}$ (according to the models LL\,\cite{tot97}, TBP\,\cite{tho02}, KRJ\,\cite{kei02}) as expected for LENA in Pyh\"asalmi after ten years of measurement. The shaded region represents the possible range of the DSNB rates depending on the normalization of the SN rate. The Super-Kamiokande limit is indicated \cite{wur07dsn}.\label{FigDsnSpe}}
\end{figure}

Fig.\,\ref{FigDsnSpe} illustrates the detection window for DSNB events in LENA: While at energies below 10\,MeV the $\bar\nu_e$ signal created by terrestrial nuclear power reactors is predominant, the $\bar\nu_e$ component of the atmospheric neutrino flux prevails above 25\,MeV. Both reactor and atmospheric background fluxes depend on the detector location. Fig.\,\ref{FigDsnSpe} presents the situation in Pyh\"asalmi.

There are a number of cosmogenic backgrounds to be taken into account: The IBD coincidence signal can be mimicked by the $\beta n$-decay of cosmogenic {$^9$Li}, fast neutrons produced either by muons in the rock and entering the detector unnoticed or by atmospheric neutrinos in NC reactions \cite{efr09}. MC calculations show that a depth of 4\,000 mwe and pulse shape discrimination of the prompt event will allow to discard these backgrounds inside a 44\,kt fiducial volume \cite{moe09dpl}.
 
The expected DSNB event rate depends on the assumptions made on the SN $\nu$ spectrum as well as on the redshift-dependent SN rate. The range of model predictions is indicated in Fig.\,\ref{FigDsnSpe} by the shaded region, lines of different color corresponding to SN models of the Lawrence-Livermore group (LL) \cite{tot97}, Thompson, Burrows and Pinto (TBP)Ê\cite{tho02}, and Keil, Raffelt and Janka (KRJ) \cite{kei02}. The detected event rate varies from about 2 to 20 events per year. Using the currently favored values for the SN rate, $\sim$10 events are expected per year \cite{wur07dsn}.

Beyond the first discovery of the DSNB signal, event rates in LENA might be sufficient to cross-check the optical measurements of red-shift dependent SN rates up to red shifts of $\sim$2. Using the input of these astronomical measurements, it might be possible to put constraints on the mean energy of the original SN $\nu$ spectrum by deconvolving the red shift. Such a measurement represents the average of the $\nu$ spectra emitted by different types of SN, and might therefore complement the observations made from a single galactic SN $\nu$ burst.

\subsection{Solar Neutrinos}

\noindent The experience with Borexino has shown that the radioactive contamination of a LSD can be reduced sufficiently to measure the solar $\nu$ spectrum down to energies of a few hundred keV \cite{bx08be7}. The spectroscopic performance of LENA will probably be inferior to the one of Borexino as the expected photoelectron yield is lower. Nevertheless, the neutrino event rates in LENA will surpass the signal in Borexino by at least two orders of magnitude. In the following, a very conservative FV of 18\,kt is chosen to ensure 7\,m of shielding against external gamma-ray background. Table\,\ref{TabLnSol} lists the expected rates for the $\nu$'s emitted in the pp chain and the CNO cycle, using the most recent solar model predictions \cite{pen08}. The rates were scaled using the results of \cite{tod08dpl}, assuming a detection threshold of 250\,keV. 

\begin{table}
\begin{center}
\begin{tabular}{|l|c|c|}
\hline
Source	& \multicolumn{2}{|c|}{Neutrino Rate [d$^{-1}$]}  \\
 			& BPS08(GS)		& BPS08(AGS)		\\
\hline
 pp 		& 24.92$\pm$0.15	& 25.21$\pm$0.13 \\
 pep 		& 365$\pm$4		& 375$\pm$4 \\
 hep 		& 0.16$\pm$0.02	& 0.17$\pm$0.03 \\
 {$^7$Be}	& 4984$\pm$297	& 4460$\pm$268 \\
 {$^8$B} 	& 82$\pm$9		& 65$\pm$7 \\
 CNO		& 545$\pm$87		& 350$\pm$52 \\
\hline
\end{tabular}
\caption{Expected solar neutrino event rates induced by neutrino-electron scattering in LENA, assuming a detection threshold of 250\,keV. Calculations are based on the high-metallicity BPS08(GS) and low-metallicity BPS08(AGS) models. The errors indicate the model uncertainties \cite{pen08}.}\label{TabLnSol}
\end{center}
\end{table}

About 25 pp-$\nu$-induced electron scattering events per day are expected above 250\,keV. It is doubtful that this rate is sufficient to be distinguished from the overwhelming {$^{14}$C} background intrinsic to the organic scintillator. About 5\,000 {$^7$Be}-$\nu$ events per day are expected: Presuming background levels and systematic uncertainties comparable to Borexino, the high statistics will allow a measurement of the {$^7$Be}-$\nu$ flux with an accuracy unprecedented in neutrino physics. It might be particularly interesting to search for temporal variations in the signal: Currently on-going MC analyses indicate that LENA will be sensitive to rate modulations of a few $\times10^{-3}$ in amplitude. In this way, fusion rate variations induced by helioseismic g-modes, $\nu$ flux modulations induced by spin flavor conversion as well as the day/night effect in terrestrial matter could be probed.

After three years of Borexino data taking it is evident that the detection of CNO and pep neutrinos decisively depends on the background level induced by $\beta$-decays of cosmogenic {$^{11}$C}. The {$^{11}$C} production rate is mainly a function of the rock overburden shielding the detector. If LENA will be operated at the intended depth of 4\,000\,mwe\,(meters water equivalent), the ratio of the CNO or pep-$\nu$ signals to the {$^{11}$C} background rate will be 1:5, a factor 5 better than in Borexino. A high-statistics measurement of about 500 CNO-$\nu$'s per day will provide valuable information on solar metallicity, especially if the contributions from the individual fusion reactions can be distinguished. The measurement of the pep-$\nu$ flux can be used for a precision test of the $\nu_e$ survival probability in the MSW-LMA transition region. Calculations indicate that also the onset of the transition region could be tested utilizing low-energetic {$^{8}$B} neutrinos \cite{tod08dpl}. 

\subsection{Geoneutrinos}

\noindent Due to the low energy threshold, LSDs are sensitive to geoneutrinos via the IBD \cite{kam05}, as has also recently been demonstrated by Borexino \cite{bx10geo}. The large target mass of LENA corresponds to roughly 1\,000 events per year if the detector is located at Pyh\"asalmi. The actual event rate is dependent on the detector location, as the geoneutrino flux depends on the crust thickness and composition near the detector site.

If the radiopurity levels of Borexino are reached in LENA, geoneutrino detection will suffer much less from internal $\alpha$-induced background than the measurements performed by KamLAND \cite{kam08}. In addition, for most of the sites investigated by LAGUNA the background from reactor neutrinos will be significantly lower. The abundances of {$^{238}$U} and {$^{232}$Th} and their natural decay chains can be determined by an analysis of the geoneutrino energy spectrum.

In spite of the high statistics, the directional information of the IBD events is not sufficient to distinguish the contributions of core, mantle, and crust to the total $\bar\nu_e$ flux: Ten years of exposure would be needed to positively identify a strong geoneutrino source of 20\,TW at the Earth's core \cite{hoc05}. A more promising approach is the combination of geoneutrino rate measurements in several LSDs at different sites: The data of KamLAND, Borexino, SNO+, LENA, and HanoHano could be combined to disentangle the geoneutrino fluxes of crust and mantle \cite{mar08}.

As about half of the Earth's thermal heat flow is unaccounted for, the occurrence of natural nuclear fission of {$^{235}$U} in the terrestrial core has been repeatedly proposed in the literature \cite{her03}: The emitted $\nu$ spectrum is assumed to be similar to the one of regular nuclear reactors which constitute the main background for detection. The current best upper limit of 3\,TW on the thermal power of this ''georeactor'' is set by the Borexino experiment \cite{bx10geo}. LENA could lower this limit to at least 2\,TW in Pyh\"asalmi, mainly because of the reduced reactor-neutrino background \cite{hoc05}.

\section{GeV Neutrino Physics}
\label{SecHigEne}

\noindent The sensitivity of LENA for the proton decay into $K^+\bar\nu$ utilizes a fast coincidence signature and is independent of the detector's tracking capabilities \cite{mar05}. However, LENA's physics potential regarding atmospheric $\nu$'s and neutrino-beam experiments is determined by the quality of the event reconstruction at GeV energies. Starting from nucleon decay, this section describes the status of the MC simulations for particle tracking \cite{lea09, pel09tra} and the expected performance of LENA as far detector in a long-baseline experiment \cite{pel09sup}.

\subsection{Nucleon Decay Search}

\noindent One of the most interesting questions in modern particle physics is the stability of the proton \cite{halzen}. While the baryon number is conserved in the standard model, most extensions predict a violation of this conservation. Grand Unified Theories usually prefer the proton decay into $\pi^0$ and $e^+$. This is also the channel water \v{C}erenkov detectors (WCDs) are able to put the most stringent limits on \cite{sk05pd}. The performance of LSDs for this decay channel heavily depends on their tracking capability at sub-GeV energies and is still to be investigated. Supersymmetry on the other hand favors the decay into $K^+$ and $\bar\nu$ \cite{bab97}. Here, WCDs have a disadvantage: Both primary decay particles are invisible in the detector as the kaon is generated at a kinetic energy below the \v{C}erenkov threshold, greatly reducing the detection efficiency.

Regarding the latter decay channel, LSDs offer a natural advantage as they are able to detect the kinetic energy deposited by the kaon \cite{mar05}. The subsequent decay into $\pi^+\pi^0$ or $\mu^+\nu_{\mu}$ provides a very fast coincidence signal ($\tau_{K^+}\simeq 13\,$ns) that can be used for background discrimination. The main background are atmospheric neutrinos in the energy regime of several 100\,MeV: Pulse shape analysis can be exploited to reduce the atmospheric background to less than 1 count in 10 years, at the same time loosing about 1/3 in proton decay sensitivity. If no signal were seen in this time span, the proton-lifetime limit could be increased to $\tau_{p}\geq 4$$\times$$10^{34}\,$yrs (at 90\,$\%$\,C.L.) for this decay mode \cite{mar05}. This surpasses the present best limit set by the Super-Kamiokande experiment by about one order of magnitude \cite{sk05pd}.

\subsection{Particle Tracking}

\begin{figure}
\centering
\includegraphics[width=0.35\textwidth]{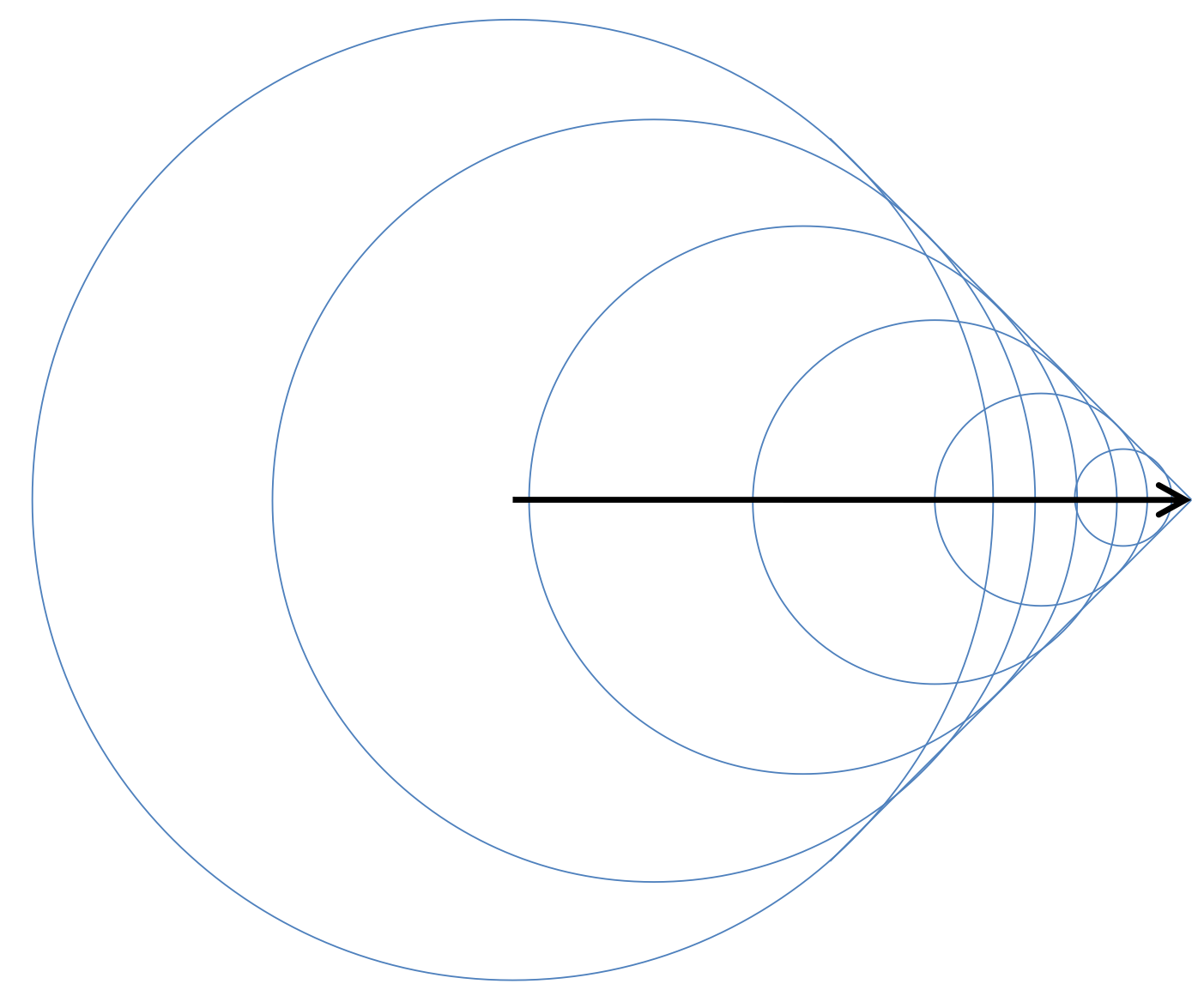}
\caption{Schematic drawing of the light front created by a muon traversing a liquid scintillator. The superposition of spherical light waves along the track creates a light front very similar to a \v{C}erenkov cone, allowing for the reconstruction of the particle direction based on PM photon arrival patterns. \label{FigLigCon}}
\end{figure}

\noindent The possibility of particle tracking in an unsegmented LSD has been neglected for a long time, although basic features are used for the reconstruction of cosmic muon tracks in the Borexino and KamLAND experiments \cite{bx08det,kam09}. However, recent studies have investigated the potential of this technique for $\nu$ event reconstruction in the GeV range. A charged particle traversing a liquid scintillator will induce scintillation along its track. At each point of the track, the produced light will be emitted isotropically. As indicated in Fig.\,\ref{FigLigCon}, the superposition of spherical light waves creates a light front which resembles the light cone typical for \v{C}erenkov light emission, adding a spherical backward running front to the v-shaped forward front. While invisible in low-energy events, the deviation from the spherical light front of point-like events is discernible for track lengths greater than the typical resolution of an LSD of tens of centimeters.

It has been demonstrated in \cite{lea09,pel09tra} that this effect can be exploited to reconstruct the tracks of charged leptons in the GeV range: The displacement of the particle vertex derived from a time-of-flight analysis of the PM hit times relative to the barycenter computed from PM hit multiplicities is already sufficient to determine the orientation of the particle track with an accuracy of a few degrees. Taking into account the total amount of collected light, particle energy and momentum can be reconstructed. The particle type is identified by the specific Bethe-Bloch ratio of visible energy to track length.

This technique allows the event reconstruction for quasi-elastic $\nu$ scattering on nucleons that dominates at energies of $\sim$1\,GeV. However, resonant single-pion production and deep-inelastic scattering prevail at higher energies. The presence of additional particles (mostly pions) considerably complicates the reconstruction of the interaction vertex. Nevertheless, the possibility to disentangle the superimposed light fronts has been demonstrated in MC simulations \cite{pel09tra}: A superposition of light patterns corresponding to MC standard events is fitted to the observed light patterns. For single-pion production, the energy of the incident neutrino can be reconstructed with an accuracy of only few per cent, including the information on the primary lepton flavor. Even for deep-inelastic interaction vertices featuring up to three pions, the lepton track and the overall event energy can be found. However, this method requires to record the signal shape of each individual PM. The signal of each PM has to be sampled by a Flash-ADC with $\leq$2\,ns time resolution for at least 100\,ns to achieve optimum results.   
 
The MC simulations presented in \cite{pel09tra} applied a simplified model of light production and propagation, are limited to a horizontal plane, and do not include the reconstruction of certain particle types, e.\,g. $\pi^0$ and neutron events. However, these first results are very encouraging, and are presently followed up by more detailed simulations based on GEANT4. This code allows a realistic reproduction of the light transport to the PMs and will be used for a systematic study of the detector's response to different particle types, including their energy and multiplicity in the interaction vertex.

\subsection{Long-Baseline Neutrino Beams}

\noindent For a small value of the mixing angle $\theta_{13}$, neutrino beam experiments searching for $\nu_e$ appearance in an accelerator-procuded $\nu_\mu$ beam are the only possibility for a precise determination of $\theta_{13}$, the CP-violating phase in the mixing matrix $\delta_\mathrm{CP}$ and the neutrino mass hierarchy. While reactor $\bar\nu_e$ disappearance experiments like the upcoming Double-Chooz experiment are sensitive to the range of $\sin^22\theta_{13}\geq0.05$\,(3$\sigma$) \cite{ard04}, the sensitivity predictions for the recently started T2K experiment reach 0.02\,(3$\sigma$) in $\sin^22\theta_{13}$ \cite{t2k01}.

Based on the current status of the tracking MC simulations, the expected performance of LENA at GeV energies \cite{pel09tra} has been used to investigate the detector's aptitude for a next-generation neutrino beam. The results of this analysis based on the GLoBES software package are reported in \cite{pel09sup}. With LENA located in Pyh\"asalmi, the distance to CERN would be 2288\,km. For this beam baseline, the neutrino energy corresponding to the first oscillation maximum is 4.2\,GeV.

The assumed neutrino source is a wide-band beam featuring energies between 1 and 6 GeV and a peak energy slightly above 1.5\,GeV. It could be produced by either the SPS or the PS2 accelerator at CERN. The projected beam power is $3.3\times10^{20}$\,POT per year or 1.5\,MW, which is about twice the power of the T2K beam. The total running time is set to 10 years, using alternating $\nu_\mu$ and $\bar\nu_\mu$ beams.

The energy resolution in LENA for CC $\nu$ events ranges from 3 to 8\,\% for energies between 1 and 5\,GeV. As indicated in Table \ref{TabHenRes}, the exact value depends on the interaction partners \cite{pel09sup}. Effects concerning the track reconstruction uncertainty, the influence of the Fermi motion and the ambiguities arising from the presence of hadronic particles, and a decrease in reconstruction efficiency above 3\,GeV were taken into account. The far detector is considered to be on-axis with respect to the beam, trading a narrower $\nu$ spectrum for larger beam intensities. In this scenario, the systematic limitations on the experimental sensitivity to a $\nu_{e}$ appearance search arise from the intrinsic contamination of the beam with $\nu_e$ ($\sim$1\,\%), while NC $\pi_{0}$ production plays only a minor role. In the $\nu_\mu$ disappearance search, the beam contamination with $\bar\nu_\mu$ as well as NC production of $\pi^\pm$ have to be considered.

\begin{table}
\begin{center}
\begin{tabular}{|c|cc|}
\hline
reaction	 & \multicolumn{2}{c|}{energy resolution}  \\
partners & at 1\,GeV		& at 5\,GeV		\\
\hline
$\bar\nu_e p$ 	& 6\,\% & 2\,\% \\
$\nu_e n$ 		& 8\,\% & 3\,\% \\
$\bar\nu_\mu p$ 	& 4\,\% & 1\,\% \\
$\nu_\mu n$ 		& 7\,\% & 2\,\%\\
\hline
\end{tabular}
\caption{The energy resolution for CC interactions of $\nu_e/\bar\nu_e$ and $\nu_\mu/\bar\nu_\mu$ in LENA. The detector performance depends on the initial neutrino energy and the interaction partners \cite{pel09sup}.}\label{TabHenRes}
\end{center}
\end{table}

The final sensitivity on oscillation parameters obviously depends on the exact values of the parameters, especially on the size of $\delta_\mathrm{CP}$. As a benchmark value, $\sin^22\theta_{13}>5\times10^{-3}$ is required to reach a $3\sigma$ discovery potential for $\theta_{13}$ and $\delta_\mathrm{CP}$, and to be able to identify the mass hierarchy. More exact values can be estimated from \cite{pel09sup}. Compared to the projected performance of T2K, this corresponds to an improvement of the sensitivity for $\theta_{13}$ by a factor 4. In addition, the CERN-Pyh\"asalmi baseline enables the search for $\delta_\mathrm{CP}$ and neutrino mass hierarchy. 

\section{Summary}
\label{SecSum}


\noindent The present contribution summarizes the status of the phenomenological work on the physics potential of LENA, a 50\,kt liquid-scintillator detector. The great potential of LSDs in the search for low-energy $\nu_e$ and especially $\bar\nu_e$ from astrophysical and terrestrial sources is undisputed. More recently, studies on the vertex resolution of GeV $\nu$ events indicate that LENA might also be a competitive option for the far detector in a long-baseline $\nu$ beam experiment. R\&D activities performed during the last years have demonstrated that an extrapolation of the design of present-day LSDs like Borexino and KamLAND to a scale of 50\,kt is technologically feasible. Suitable candidates for the liquid scintillator have been identified in extensive laboratory tests, favoring LAB at present. The construction of the underground laboratory and detector cavern as well as the tank design is currently being investigated within the LAGUNA site study. The optimization of the PM configuration and the design of the read-out electronics have already started. A realization of the detector within the upcoming decade seems feasible. LENA might take its first low-energy neutrino data before the year 2020. 

\section*{Acknowledgements}
\noindent This work has been supported by the Maier-Leibnitz-Laboratorium (Gar\-ching), the Deutsche Forschungsgemeinschaft DFG (Transregio 27: Neutrinos and Beyond), and the Munich Cluster of Excellence (Origin and Structure of the Universe).

\bibliographystyle{h-physrev}
\bibliography{epiphany_arxiv}

\end{document}